\documentclass{article}
\usepackage[preprint]{spconf}
\usepackage{amsmath,graphicx}
\usepackage{ amssymb }
\usepackage{algorithm2e}
\usepackage{multirow}
\usepackage{flushend}
\onecolumn

\title{ITERATIVE CAUCHY THRESHOLDING:\\ REGULARISATION WITH A 
HEAVY-TAILED PRIOR}
\name{Perla Mayo$^{\dagger}$\thanks{\(^{1}\)This work was supported in part by a CONACyT PhD studentship under grant 461322 (to Mayo), in part by the Engineering and Physical Sciences Research Council (EPSRC) under grant EP/R009260/1 (AssenSAR), and in part by a Leverhulme Trust Research Fellowship (to Achim).}, Robin Holmes$^{\ddag}$ and Alin Achim$^{\dagger}$}
\address{$^{\dagger}$ Visual Information Lab, University of Bristol, Bristol, U.K. \\
$^{\ddag}$University Hospitals Bristol NHS Foundation Trust, Bristol, U.K.}
\begin{document}
%
\maketitle
\begin{abstract}
In the machine learning era, sparsity continues to attract significant interest due to the benefits it provides to learning models. 
Algorithms aiming to optimise the \(\ell_0\)- and \(\ell_1\)-norm are the common choices to achieve sparsity. In this work, an alternative algorithm is proposed, which is derived based on the assumption of a Cauchy distribution characterising the coefficients in sparse domains. The Cauchy distribution is known to be able to capture heavy-tails in the data, which are linked to sparse processes. We begin by deriving the Cauchy proximal operator and subsequently propose an algorithm for optimising a cost function which includes a Cauchy penalty term. We have coined our contribution as Iterative Cauchy Thresholding (ICT). Results indicate that sparser solutions can be achieved using ICT in conjunction with a fixed over-complete discrete cosine transform dictionary under a sparse coding methodology.
\end{abstract}
\section{Introduction}
\label{sec:intro}
Representation learning (RL) involves the acquisition of features that enable the reconstruction or recovery of a signal of interest. Such features can be learnt in several ways. In dictionary learning it is assumed that an observed signal \(\textbf{y} \in \textrm{I\!R}^{M}\) can be approximately represented or reconstructed by a linear combination of features in a matrix \(\textbf{A} \in \textrm{I\!R}^{MxN}\) with a vector of coefficients \(\textbf{x} \in \textrm{I\!R}^{N}\). This is expressed as:

\begin{equation}
    \label{eq:y_hat}
    \hat{\textbf{y}} = \textbf{Ax}
\end{equation}

\noindent 
where \(\hat{\textbf{y}} \in \textrm{I\!R}^{M}\) is the estimation of the observed signal \(\textbf{y}\). The error between this approximation and the true signal is thus given by \(\mathcal{L}(\textbf{A}, \textbf{x}) = || \textbf{y} - \hat{\textbf{y}}||_2^2\). In Eq. (\ref{eq:y_hat}), each coefficient \(x_i\) provides the contribution of a column vector (feature) \(\textbf{a}_i\) in the reconstruction task for \(i=1,..., N\). Additionally, along with learning of the vectors in \textbf{A}, it is of interest to understand their contribution. The goal is then to recover the coefficients within \(\textbf{x}\), which is done through the optimisation of 

\begin{equation}
    \label{eq:opt_noreg}
    \begin{aligned}
        \textbf{x}^* &= \underset{\textbf{x}}{\text{argmin}}
		& & \mathcal{L}(\textbf{A}, \textbf{x})  
		             = \underset{\textbf{x}}{\text{argmin}}
		& & ||\textbf{y} - \textbf{Ax}||_{2}^{2}  \\
	\end{aligned}
\end{equation}

When the matrix \(\textbf{A}\) is over-complete (\(N >> M\)), the problem in Eq. (\ref{eq:opt_noreg}) becomes ill-posed with an infinite number of solutions. To alleviate this, a regularisation term is added in Eq. (\ref{eq:opt_noreg}) to enforce sparsity among the elements of the coefficients \(x_i\):

\begin{equation}
    \label{eq:opt_reg}
    \begin{aligned}
		\textbf{x}^* &= \underset{\textbf{x}}{\text{argmin}}
		& & \mathcal{L}(\textbf{A, x}) + \lambda \varphi (\textbf{x}) \\
	\end{aligned}
\end{equation}

\noindent
where \(\lambda\) is a trade-off parameter between data fidelity and the penalty term.

In Eq. (\ref{eq:opt_reg}), the \(\ell_0\)-norm is the desired choice for \(\varphi(\cdot)\). This norm is ideal since it effectively counts the number of non-zero entries in \(\textbf{x}\), therefore, the problem is to find the optimal set of coefficients \(x_i\) that will allow the reconstruction of \(\textbf{y}\) by using as few as possible column vectors from \(\textbf{A}\). Since this is a combinatorial (NP-hard) problem, its optimisation becomes challenging. An alternative to this is to replace the \(\ell_0\)- by its relaxed version, the \(\ell_1\)-norm. Furthermore, different penalty functions can be obtained if a different approach is followed. When the coefficients are assumed to follow a heavy-tailed distribution centred at 0 it is possible to derive alternative penalties. 
If a Laplace distribution is assumed for the coefficients, then the penalty function corresponds to the \(\ell_1\)-norm \cite{grosse2012sisc}. 
The use of heavy tailed distributions has also been considered, e.g. in \cite{pad2017sparsedt}, where it was assumed that coefficients follow the Symmetric \(\alpha\)-Stable distribution. 
Our aim is to prove the power of the Cauchy distribution in modelling data with heavy tails, and consequently design novel sparse coding approaches.

The rest of the paper is structured as follows: in section \ref{sec:related} a brief summary of related work is presented along with the motivation behind it, section \ref{sec:proposed} describes the proposed algorithm in detail, section \ref{sec:experiments} presents the experiments carried out along with the obtained results. Lastly, a conclusion is presented in section \ref{sec:conclusions} along with a discussion of future work.

\section{Related Work}
\label{sec:related}
The optimisation of either the \(\ell_0\)- or \(\ell_1\)-norms can be done through iterative thresholding approaches such as, e.g., the Iterative Hard Thresholding (IHT) \cite{blumensath2008ihtsparse} and the Iterative Soft Thresholding (IST) \cite{beck2009fista} algorithms. They are expressed as shown in equations (\ref{eq:iht}) and (\ref{eq:ist}) respectively.

\begin{equation}
    \label{eq:iht}
    x = 
    \begin{cases}
         x, & \text{if } |x| > x_0 \\
         0, & \text{if } |x| \leq x_0 
    \end{cases}
\end{equation}

\begin{equation}
    \label{eq:ist}
    x = 
    \begin{cases}
         x - x_0, & \text{if } x > x_0 \\
         x + x_0, & \text{if } x < -x_0 \\
         0, & \text{otherwise}
    \end{cases}
\end{equation}

These approaches achieve sparsity by snapping to zero any value below a given threshold. In the case of IST the values are shrunk by \(x_0\). In addition, the function in (\ref{eq:ist}) can be derived by computing the proximal operator of the penalty function, obtained by solving the expression in Eq. (\ref{eq:proxop}) below.

\begin{equation}
    \label{eq:proxop}
    \begin{aligned}
		\text{prox}_{\lambda\varphi}(x) &= \underset{\textbf{z}}{\text{argmin}}
		& & (z-x)^2 + \lambda \varphi (z) \\
	\end{aligned}
\end{equation}

 For IST, using \(\varphi(\cdot) = |\cdot|\) results in the expression in (\ref{eq:ist}) with \(x_0 = \frac{\lambda}{2}\). For more details on proximal algorithms, please refer to \cite{parikh2014proximalalgorithms, combettes2011proximal}.

\subsection{Cauchy distribution as prior for sparsity}
\label{subsec:cauchyprior}
As previously said, the penalty function \(\varphi(\cdot)\) can be defined from a maximum a posteriori (MAP) approach by assuming the coefficients in \(\textbf{x}\) follow a heavy tailed distribution centred at 0. The Laplacian distribution is implicitly chosen when the regularisation term is the \(\ell_1\)-norm. The Cauchy distribution has been found to be an interesting alternative~\cite{Wan11_iet-ip}. It belongs to the family of the \(\alpha\)-Stable distributions and it is one of the few that has a closed form expression for its probability density function (p.d.f.), which is defined in Eq. (\ref{eq:cauchypdf}).

\begin{equation}
    \label{eq:cauchypdf}
    p(x) = \frac{\gamma}{\pi(\gamma^2 + (x-\delta)^2)}
\end{equation}

\begin{figure}
    \label{fig:cauchypdfs}
    \centering
    \includegraphics[scale=0.5]{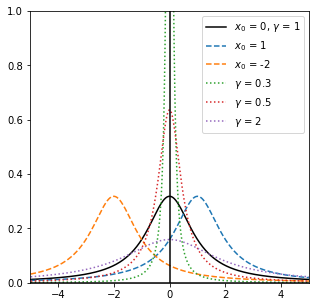}
    \caption{Cauchy PDF with different parameters}
\end{figure}

\noindent
The Cauchy distribution is described by two parameters for location (\(\delta\)) and scale (\(\gamma\)). The behaviour of these parameters is illustrated in Fig. \ref{fig:cauchypdfs}. It can be seen that a smaller value for \(\gamma\) corresponds to sparser densities and a larger concentration around the location  \(\delta\). When \(\delta=0\), most values will concentrate around the origin. In addition, the smaller \(\gamma\), the more aggressive the sparsity enforcing property of the corresponding algorithm will be.

\section{A new shrinkage algorithm}
\label{sec:proposed}
The shrinkage algorithm proposed in this work relies on the proximal splitting approach for a Cauchy penalty function. This results in the optimisation of (\ref{eq:opt_reg}) one function at a time. Furthermore, \(\varphi(\cdot)\) is solved using the shrinkage operator obtained from the computing of the proximal operator for the Cauchy penalty function. This is further reviewed in section \ref{subsec:proxop_cauchy} whilst the full algorithm is described in detail in section \ref{subsec:ict}.

\subsection{The Cauchy Proximal Operator}
\label{subsec:proxop_cauchy}
The Cauchy proximal operator is obtained when Eq. (\ref{eq:proxop}) is solved using \(\varphi(x) = -\text{log }p(x)\) (from Eq. (\ref{eq:cauchypdf})) and assuming \(\delta = 0\). Therefore, the proximal operator for this penalty function is computed from:

\begin{equation}
    \label{eq:cauchy_proxop}
    \begin{split}
        \text{prox}_{\lambda \varphi}(x) &= \text{argmin   } (z - x)^2 + \lambda \varphi(z) \\
         &= \text{argmin   } (z - x)^2 - \lambda\text{log}\left(\frac{\gamma}{\pi \left(\gamma^2 + z^2 \right) }\right)
    \end{split}
\end{equation}

\noindent
up to a constant, which results in:

\begin{equation}
    \label{eq:ict_min}
    z^3 - x z^2 + (\gamma^2+ \lambda) z - \gamma^2 x = 0
\end{equation}

\noindent
Using Cardano's formula to solve this cubic expression, we obtain:

\begin{equation}
    \label{eq:cauchy_shrinkage}
    z = \frac{x}{3} + t
\end{equation}

\noindent
where
\begin{equation*}
    \begin{split}
        t &= \sqrt[3]{-\frac{q}{2}+\sqrt[2]{\Delta}} \\
        \Delta &= \frac{q^2}{4}+\frac{p^3}{27} \\
        p &= \lambda + \gamma^2-\frac{x^2}{3} \\
        q &= -\frac{2}{27}x^3 + \frac{1}{3}\left(\lambda-2\gamma^2\right)x
    \end{split}
\end{equation*}

This will provide only one real root when \(\Delta > 0\), and three real roots otherwise. In the latter case the root with the largest absolute value is chosen. 

The shrinkage behaviour offered by the Cauchy proximal operator does not require one to explicitly specify a threshold value, as must be done for IST and IHT. However, the shape of the Cauchy proximal operator still depends on the parameters selection. Such shape is an indicator of how aggressive the cut-off will be. Figure \ref{fig:cauchy_behaviour}(a) shows the impact of the parameter \(\lambda\) in the Cauchy proximal operator, whilst Figure \ref{fig:cauchy_behaviour}(b) does the same for \(\gamma\). It can be seen that \(\lambda\) determines the location of the threshold and \(\gamma\) its shape. A more aggressive cut-off will be exhibited when \(\gamma \rightarrow 0\).

In fact, when \(\gamma \rightarrow 0 \), its value will be so small that it will barely contribute anything to the function to minimise. If \(\gamma = 0\), the function to minimise becomes:

\begin{figure}
    \begin{minipage}[b]{.48\linewidth}
        \centering
        \centerline{\includegraphics[width=3.7cm]{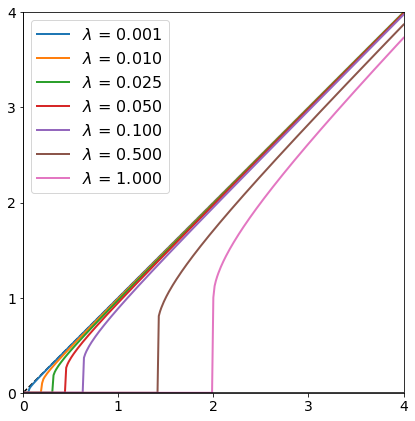}}
        \centerline{(a)}\medskip
    \end{minipage}
    \hfill
    \begin{minipage}[b]{0.48\linewidth}
        \centering
        \centerline{\includegraphics[width=3.7cm]{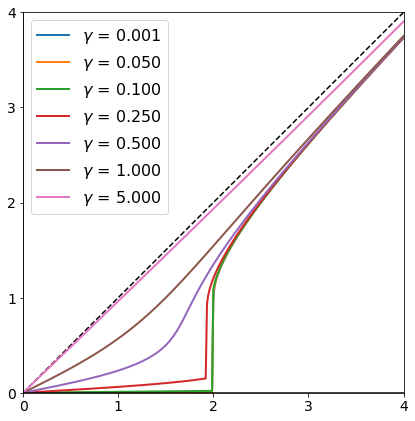}}
        \centerline{(b)}\medskip
    \end{minipage}
    \caption{Behaviour of the Cauchy Thresholding algorithm with a fixed value of a) \(\gamma = 0.001\), b) \(\lambda = 1\).}
    \label{fig:cauchy_behaviour}
\end{figure}

\begin{equation*}
    \begin{split}
        \text{prox}_{\lambda \varphi}(x) 
         &= \text{argmin   } (z - x)^2 + \lambda\text{log}(z^2)
    \end{split}
\end{equation*}

\noindent
which is now a non-smooth non-convex function. Its solutions, however, can be computed following the proximal operator approach once again:

\begin{equation*}
    z = 
    \begin{cases}
         \frac{x}{2} + \frac{\sqrt{x^2-4\lambda}}{2} &, x \geq 2\lambda\\
         \frac{x}{2} - \frac{\sqrt{x^2-4\lambda}}{2} &, x \leq -2\lambda\\
         0, & \text{otherwise}
    \end{cases}
\end{equation*}

The shape of ICT under these conditions is shown in figure \ref{fig:cauchy_behaviour}b): as \(\gamma \rightarrow 0\), the threshold \(\rightarrow 2\) (\(\lambda = 1\)). On the other hand, when \(\lambda = 0\), the solutions are given by \(\gamma i, -\gamma i\), and \(x\), which implies that there is no shrinkage. This comes from the fact that \(\lambda=0\) removes the penalty on the coefficients from \(\textbf{x}\), and so the learning relies purely on the reconstruction error.

\begin{figure}
    \centering
    \includegraphics[scale=.4]{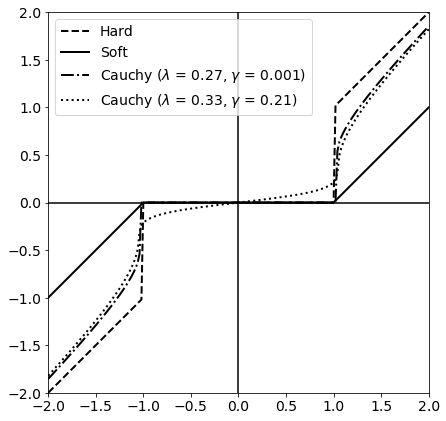}
    \caption{Comparison of different thresholding algorithms}
    \label{fig:diff_thres}
\end{figure}

\subsection{Iterative Cauchy Thresholding}
\label{subsec:ict}
Once the proximal operator has been obtained, it is our goal to solve Eq. (\ref{eq:opt_reg}) for the coefficients using the Cauchy penalty function. Explicitly, the cost function is now:

\begin{equation}
    \label{eq:cauchy_cost}
    f(\textbf{x}) = ||\textbf{y}-\textbf{Ax}||_2^2 - \sum_{i}log\left(\frac{\gamma}{\pi(\gamma^2 + x_i^2)}\right)
\end{equation}
\begin{table}[t]
    \centering
    \footnotesize
    \begin{tabular}{|c|c|c|c|c|}
        \hline
        \textbf{Dataset} & \textbf{Metric} & \textbf{IHT} & \textbf{IST} & \textbf{ICT} \\
        \hline
        \multirow{2}{*}{MNIST} 
            & PSNR & 50.70 & 50.68 & 50.70\\
            & \% Non-zeros & 85.07 & 85.06 & 83.83\\
        \hline
       \multirow{2}{*}{AT\&T Faces} 
            & PSNR & 47.39 & 47.38 & 47.31\\
            & \% Non-zeros & 98.14 & 99.99 & 75.6 \\
        \hline
        \multirow{2}{*}{Lena} 
            & PSNR & 47.26 & 47.26 & 47.26 \\
            & \% Non-zeros & 99.80 & 99.98 & 95.18 \\
        \hline
        \multirow{2}{*}{Shepp-Logan} 
            & PSNR & 53.77 & 53.76 & 53.77 \\
            & \% Non-zeros & 41.99 & 44.74 & 41.12 \\
        \hline
    \end{tabular}
    \caption{Best PSNR values for reconstruction using the different thresholding algorithms after 200 iterations}
    \label{tab:results_psnr}
\end{table}

\begin{table}[t]
    \centering
    \footnotesize
    \begin{tabular}{|c|c|c|c|c|}
        \hline
        \textbf{Dataset} & \textbf{Metric} & \textbf{IHT} & \textbf{IST} & \textbf{ICT} \\
        \hline
        \multirow{2}{*}{MNIST} 
            & PSNR & 19.52 & 13.95 &  19.50\\
            & \% Non-zeros & 3.87 & 2.62 & 3.87\\
        \hline
       \multirow{2}{*}{AT\&T Faces} 
            & PSNR & 24.02 & 13.59 & 23.99\\
            & \% Non-zeros & 4.12 & 1.66 & 4.11 \\
        \hline
        \multirow{2}{*}{Lena} 
            & PSNR & 26.92 & 13.67 & 26.87 \\
            & \% Non-zeros & 4.08 & 1.64 & 4.07 \\
        \hline
        \multirow{2}{*}{Shepp-Logan} 
            & PSNR & 24.61 & 17.78 & 14.07 \\
            & \% Non-zeros &  1.40 & 0.72 & 0.03 \\
        \hline
    \end{tabular}
    \caption{Sparsest results for image reconstruction using the different thresholding algorithms after 200 iterations.}
    \label{tab:results_sparsest}
\end{table}

This is solved by means of a proximal splitting approach. Under this framework, the function \(f(\textbf{x})=\mathcal{L}(\textbf{x}) + \lambda\varphi(\textbf{x})\) can be optimised by separating \(\mathcal{L}(\textbf{x})\) and \(\lambda\varphi(\textbf{x})\), therefore, the solutions for each function are projected into the next one. For this, only the gradient of \(\mathcal{L}(\textbf{x})\) is computed to solve \(\mathcal{L}(\textbf{x})\). Such solution is then passed to the solver for the Cauchy penalty function, which is the shrinkage function given by Eq. (\ref{eq:cauchy_shrinkage}). The shrinkage is applied in an entry-wise fashion. The process is iterated until a stopping criterion is met, that is, after a fixed number of iterations or when the error has reached a certain predefined value. This is the Iterative Cauchy Thresholding (ICT) algorithm, and its pseudo-code is shown in Algorithm \ref{alg:ict}.

\begin{algorithm}
    \SetAlgoLined
    \KwResult{Sparse coefficients \(\textbf{x}\)}
    Initialise \(\textbf{x}\) with 0's\;
    Set \(\eta\), \(\gamma\) and \(\lambda\)\;
    Choose stopping criteria\;
    \While{Stopping criteria has not been met}{
        Compute \(\textbf{x} = \textbf{x} - \eta\nabla_\textbf{x}\mathcal{L}(\textbf{A}, \textbf{x})\)\;
        Shrink \(\textbf{x}\) using Eq. (\ref{eq:cauchy_shrinkage}).
    }
    \caption{Iterative Cauchy Thresholding}
    \label{alg:ict}
\end{algorithm}

\section{Experimental Results}
\label{sec:experiments}
The performance of ICT has been assessed for a reconstruction task via Sparse Coding (SC). The problem to solve for SC is the same as the one posed in Eq. (\ref{eq:opt_reg}). The performance of ICT is compared against the ones for IST and IHT, where only the penalty function changes during the optimisation accordingly. The dictionary used is the fixed over-complete discrete cosine transform (DCT) using a patch size of 8x8 and a total of 144 atoms. Due to the nature of SC, the image to reconstruct is thus split in \(T\) overlapping patches of the same size as the atoms (8x8). There is one vector of sparse coefficients \(\textbf{x}\) for every one of these patches. The reconstructed image corresponds to the average of the reconstructed overlapping patches. Further details on sparse coding are beyond the scope of this work and can be found elsewhere \cite{olshausen1997sparsev1,lee2007efficientsparsecoding}.

\begin{figure}[t]
    \begin{minipage}[b]{.48\linewidth}
        \centering
        \centerline{\includegraphics[width=4.0cm]{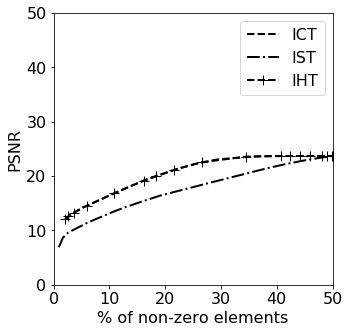}}
        \centerline{(a)}\medskip
    \end{minipage}
    \hfill
    \begin{minipage}[b]{0.48\linewidth}
        \centering
        \centerline{\includegraphics[width=4.0cm]{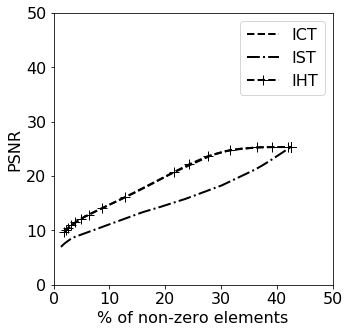}}
        \centerline{(b)}\medskip
    \end{minipage}
    \begin{minipage}[b]{.48\linewidth}
        \centering
        \centerline{\includegraphics[width=4.0cm]{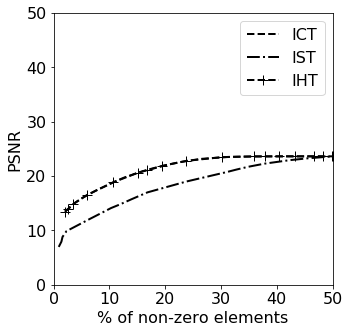}}
        \centerline{(c)}\medskip
    \end{minipage}
    \hfill
    \begin{minipage}[b]{0.48\linewidth}
        \centering
        \centerline{\includegraphics[width=4.0cm]{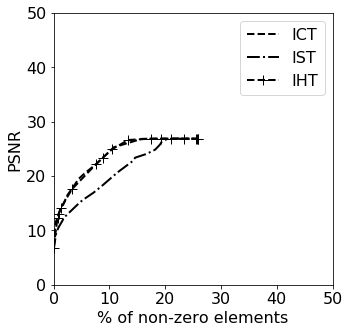}}
        \centerline{(d)}\medskip
    \end{minipage}
    \caption{Average of \% of non-zero elements in the coefficients vs PSNR for the three thresholding algorithms for the different datasets: (a) AT\&T Faces, (b) MNIST, (C) Lena, and (4) Shepp-Logan.}
    \label{fig:sparsity_vs_psnr}
\end{figure}

The images to reconstruct correspond to the AT\&T faces dataset \footnote{www.cl.cam.ac.uk/research/dtg/attarchive/facedatabase.html}, the MNIST\footnote{http://yann.lecun.com/exdb/mnist/} dataset as well as the Lena and the Shepp-Logan phantom images. The thresholding algorithms were evaluated with varying values for their parameters. To obtain different levels of sparsity the parameter \(\lambda\) was modified for the three algorithms. 
The learning rate \(\eta\) was fixed for the three algorithms to 0.005. We randomly chose 500 observations for the MNIST dataset and 50 for the AT\&T faces. 
Tables \ref{tab:results_psnr} and \ref{tab:results_sparsest} report the average results on this reconstruction task for the three thresholding algorithms, PSNR values along with the \% of non-zero elements (\(\ell_0\)-norm) are reported, whilst Figure \ref{fig:sparsity_vs_psnr} shows the performance as function of the iterations for the four datasets.

The performance of ICT is very similar to the one produced by IHT, both achieving better PSNRs than the  reconstruction results obtained with IST by a considerable margin, for most levels of sparsity considered. The best PSNR values obtained by the three algorithms correspond to the lowest degree of sparsity for each and vice-versa. It is the latter case the one of interest. 
The similarity in behaviour between IHT and ICT could be attributed to the shape of their functions. In Fig. \ref{fig:diff_thres} it can be seen that the ICT shrinkage function lies between those of both IHT and IST and it affects all coefficients regardless of their amplitude.
It is important to note as well that the reconstruction itself corresponds to the average of all estimations obtained for every overlapping patch extracted from the original observations. Hence there will always be a subtle blurring effect affecting the final image. This can be alleviated by employing a different framework, such as convolutional sparse coding \cite{grosse2012sisc, zeiler2010deconvolutional}.

\section{Conclusions and Future Work}
\label{sec:conclusions}
 In this paper we introduced a new algorithm to recover sparse coefficients in regularised inverse problems. The proposed approach is based on the assumption that sparse coefficients can be accurately  modelled by Cauchy distributions. Consequently, we introduce the Cauchy proximal operator and describe a corresponding proximal splitting approach, which we prefer to refer to as ICT. As an illustrative example, we chose to address the problem of image reconstruction from learnt sparse representations, and we achieve a performance which compares very favourably with the IHT algorithm. We conjecture that the proposed ICT algorithm has the potential to perform better in other inverse problems, including, e.g. image denoising or deconvolution, due to the optimal penalty applied to large coefficients. This is the focus of our current research endeavours.

\bibliographystyle{IEEEbib}
\bibliography{main}

\end{document}